\documentclass[aps,prl,twocolumn,nofootinbib,amssymb,amsfonts,superscriptaddress,longbibliography]{revtex4-1}
\usepackage{amssymb}
\usepackage{color}
\usepackage{graphicx}
\usepackage{epsfig,amssymb,amsmath,amsthm}

\usepackage[utf8]{inputenc} 
\usepackage[T1]{fontenc}

\def\>{\rangle}
\def\<{\langle}

\def\labell#1{\label{#1}}
\def\togli#1{}

\begin{document}

\title{The time-energy uncertainty relation for quantum events}

\author{Matteo Fadel}
\email{fadelm@phys.ethz.ch} 
\affiliation{Department of Physics, ETH Z\"urich, 8093 Z\"urich, Switzerland} 
\affiliation{Department of Physics, University of Basel, Klingelbergstrasse 82, 4056 Basel, Switzerland}

\author{Lorenzo Maccone}
\affiliation{Dip.~Fisica and INFN Sez. Pavia, University of Pavia, via Bassi 6, I-27100 Pavia, Italy}

\begin{abstract}
    Textbook quantum mechanics treats time as a classical parameter, and not as a quantum observable with an associated Hermitian operator. For this reason, to make sense of usual time-energy uncertainty relations such as $\Delta {t}\Delta E\gtrsim\hbar$, the term $\Delta t$ must be interpreted as a time interval, and not as a time measurement uncertainty due to quantum noise. However, quantum clocks allow for a measurement of the ``time at which an event happens'' by conditioning the system's evolution on an additional quantum degree of freedom. Within this framework, we derive here two {\em true} uncertainty relations that relate the uncertainty in the quantum measurement of the time at which a quantum event happens on a system to its energy uncertainty.
\end{abstract}

\maketitle

While classical mechanics allows for the simultaneous measurement with
arbitrary precision of any set of physical observables, quantum
mechanics predicts situations in which this is fundamentally
forbidden. Concrete examples include position/momentum and different
angular momentum or spin components.

For physical observables associated to Hermitian operators,
uncertainty relations provide quantitative lower bounds on their
simultaneous measurement uncertainty. The Heisenberg-Robertson
\cite{robertson,trifonov} uncertainty relation
$\Delta A\Delta B\geqslant|\<[A,B]\>|/2$ expresses a tradeoff between
the root-mean-square-error (RMSE) $\Delta X = \sqrt{\<X^2\>-\<X\>^2}$
of two observables $A$ and $B$ in terms of their commutator. By
considering also the anticommutator, a tighter version of this
inequality was found by Schr\"odinger \cite{sch} to be
$\Delta A^2\Delta
B^2\geqslant|\<[A,B]\>/2|^2+|\<\{A,B\}_+\>/2-\<A\>\<B\>|^2$. In both
these relations, uncertainties originate from the stochastic nature of
quantum measurements, and the bound from quantum complementarity: it
is impossible to know simultaneously incompatible observables with
arbitrary precision.

It is tempting, but wrong, to give a similar interpretation to
uncertainty relations between time and energy \cite{busch}. In fact,
in textbook quantum mechanics time is not a quantum observable, but a
classical parameter with no intrinsic quantum uncertainty. For this
reason, any term $\Delta t$ has to be understood as a time interval.
For example, the correct interpretation of the Mandelstamm-Tamm
relation \cite{man} is: the smallest time {\em interval} $\Delta t$
required for a system with energy spread $\Delta E$ to evolve into an
orthogonal state is lower bounded by $\Delta E\Delta {t}\gtrsim\hbar$.
This interpretation follows immediately from the (classical)
inequality between time {\em duration} $\Delta t$ and bandwidth
$\Delta\omega$ of a signal, $\Delta{t} \Delta\omega\gtrsim 1$,
together with Planck's relation $E=\hbar\omega$. Interestingly, a
similar bound was also given in terms of the average energy by
Margolus and Levitin \cite{margolus}: the smallest time interval
required for a system with average energy $\<E\>$ (above the ground
state) to evolve into an orthogonal state is lower bounded by
$\Delta {t}\geqslant\pi\hbar/2\<E\>$. Both these relations can also be
extended to the case in which the evolved state has arbitrary overlap
with the initial state \cite{bata,pfeiffer,qlimits} In essence, rather
than true uncertainty relations, these inequalities are more properly
``quantum speed limits'' \cite{davidov,steve}, bounding the ``speed''
of a dynamical evolution.  Other uncertainty relations assign to
$\Delta t$ the minimum time {\em interval} required to estimate the
energy of a system with precision $\Delta E$.  While this last
interpretation is in general incorrect \cite{aharonov}, it is valid if
the system's Hamiltonian is unknown \cite{popescu,cohen}. Therefore,
in all the aforementioned relations (Mandelstamm-Tamm,
Margolus-Levitin, Aharonov-Massar-Popescu) the quantity $\Delta t$ is
never a RMSE due to quantum noise, but rather a time interval.
Energy--time uncertainties also arise in quantum gravity, for clocks
of limited dimensions \cite{bronstein}: a high precision of the clock
requires a large energy devoted to it, but if that energy is
compressed inside its Schwarzschild radius, a gravitational collapse
occurs and the clock ceases to keep time. Similar effects are due to
time dilation \cite{gambini}. All these arguments were not obtained
from quantum complementarity of observables.

In this paper we prove a {\em true} time--energy uncertainty relation
\begin{align}
\Delta {t}_{ev}\Delta E_{ev}\geqslant {\hbar}/{2}
\labell{teunc}\;,
\end{align}
where ${t}_{ev}$ is the time at which some event happens in a quantum system and $E_{ev}$ is the system's energy conditioned on the event happening. In contrast to the time--energy uncertainty relations mentioned before, the quantity $\Delta {t}_{ev}$ refers now to the uncertainty (RMSE) in the measurement of a quantum time observable due to quantum noise. The Hermitian time operator ${T}_c$, from which $t_{ev}$ can be obtained through the Born rule, is constructed here by considering an ancillary quantum system that serves as a clock \cite{arrival}. 
Crucially, the energy of the system conditioned on the event happening, $E_{ev}$, can be connected to a system observable only if the projector $\Pi$ that tests whether the even has happened is compatible (i.e. commutes) with the system's Hamiltonian $H_s$. In this case $E_{ev}=\Pi H_s\Pi$. Otherwise, the ``conditioned energy'' is not even a well defined concept due to quantum complementarity.  Nonetheless, even if $[\Pi,H_s]\neq 0$, we show that one can still introduce such a quantity thanks to a constraint equation between the system's and the clock's energies, which allows to condition on the clock's energy that is always compatible with $\Pi$.

Time--energy relations considering the RMSE of a time measurements
have also been introduced in \cite{seth,ahmadi,verrucchi}. These,
however, refer only to uncertainties in the measurement of time (or
proper time) with a clock of given energy, rather than the measurement
of the {\em time at which an event happens}, as we do here.

\textit{Quantum time measurement.--} To give a quantum description of a time measurement, we first need to treat the clock that is used for this purpose as a quantum system \cite{paw,aharonovt,qtime}. Then, the quantum time measurement is obtained from the observable ${T}_c\equiv\int dt\:{t}\:|{t}\>\<{t}|$, with $|t\>$ the clock state associated to what happens to the system at time $t$ \cite{arrival}. To see this, let us define the timeless state $|\Psi\>\>$ that contains the full dynamics of the system's state
$|\psi({t})\>$ by correlating it to the clock's state $|t\>$ as
\begin{align}
|\Psi\>\>=\frac 1{\sqrt{{T}}}\int_{-{T}/2}^{{T}/2}dt\:|{t}\>|\psi({t})\>
\labell{psipp}\;.
\end{align}
Here, the double ket notation is just a reminder that $|\Psi\>\>$ is a joint clock--system state, and $T$ is a regularization parameter that represents the total time interval we consider (${T}\to\infty$ considers the full evolution of the system).

The state $|\Psi\>\>$ is an eigenstate of the Dirac-type \cite{dirac} constraint operator $(H_c\otimes\openone_s+\openone_c\otimes H_s)|\Psi\>\>=0$, where $H_{s(c)}$ is the system (clock) Hamiltonian. In order for $|\psi({t})\>$ to satisfy the Schr\"odinger equation, $H_c$ must coincide with the clock's momentum, namely $[{T}_c,H_c]=i\hbar$ \cite{paw,qtime}. The time observable ${T}_c$ can thus be interpreted as the ``position'' observable for the clock, conjugate to its energy (such that the clock's energy is the generator of time translations).

By ``the event happens'' one means that some property of the system
acquires a value that is connected with the event. E.g.~if one wants
to measure ``the time at which the spin is up'', then one must measure
the time at which the spin value is ``up''. Each value of a system
property, identified by some system observable, is connected to a
projector $\Pi$. It refers to the observable eigenspace relative to
the eigenvalue(s) connected to the event's value for the property.
E.g.~for the above example, $\Pi={|\!\uparrow\>\<\uparrow\!|}$ is the
projector on the system's ``spin-up'' state.

Consider the observable ${T}_\pi={T}_c\otimes\Pi$, where $\Pi$ is the
projector relative to the value of the property that indicates that
the event happens.  The Born rule tells us that the {\em joint}
probability that the clock shows time $t$ {\em and} that the event has
happened is
\begin{align}
  p({t},\Pi)=
  \mbox{Tr}[|{t}\>\<{t}|\otimes\Pi|\Psi\>\>\<\<\Psi|]
\labell{pborn}\;,
\end{align}
(e.g. $p({t},\Pi)=\left|\<\<\Psi|{t}\>|\!\uparrow\>\right|^2$ for the ``spin-up event'' example).
Similarly, the probability that the event happens at any time is $p(\Pi)=\<\<\Psi|\Pi|\Psi\>\>=P_{ev}$, which follows from $\int dt|{t}\>\<{t}|=\openone$. From these two probabilities, using Bayes' rule, we can calculate the conditional probability that the clock shows time $t$ given that the event happened as $p({t}|\Pi)=p({t},\Pi)/p(\Pi)$. This probability allows us to compute the expectation value for the time at which the event happened as
\begin{align}
\<{t}_{ev}\>=\int dt\:{t}\:p({t}|\Pi)=\dfrac{\<{T}_\pi\>}{\<\Pi\>} =\alpha {T}\<{T}_\pi\>
\labell{avgt}\;,
\end{align}
where all expectation values are calculated on $|\Psi\>\>$ and $\alpha^{-1}\equiv\int dt \<\psi({t})|\Pi|\psi({t})\>=\<\Pi\>T$, namely $(\alpha {T})^{-1}=p(\Pi)$ is the probability that the event happened at any time. 
The same conditional probability distribution allows us to calculate also the variance \footnote{This formula is written incorrectly in \cite{arrival}, where the $T$ factors are missing.}
\begin{align}
\Delta {t}^2_{ev}=\<{t}_{ev}^2\>-\<{t}_{ev}\>^2=\alpha
{T}\<{T}_\pi^2\>-\alpha^2T^2\<{T}_\pi\>^2 \;,
\labell{var}\;
\end{align}
which expresses the uncertainty in the measurement of the time $t_{ev}$ at which the event happens. Whenever $\<t_{ev}\>=0$ it is clear that $\<T_\pi\>=0$, which implies $\Delta t_{ev}^2=\alpha {T}\Delta T_\pi^2$. The latter equality, however, is true also in all other cases where $\<t_{ev}\>\neq 0$, since a shift of the averages will not affect the variances, and a shift ${t}_0$ of the average value of $t_{ev}$ corresponds to a shift $\alpha {T} {t}_0$ of the average value of $T_\pi$, since $\<t_{ev}\>=\<T_\pi\>\alpha {T}$. In conclusion, we always have $\Delta {t}_{ev}^2=\alpha T\Delta {T}_\pi^2$.

 \textit{Conditional energy.--} In order to prove
Eq.~\eqref{teunc}, we now need to evaluate $\Delta H_{ev}$, namely the
RMSE of the energy conditioned on the event having happened. This term
can be calculated from the clock energy, thanks to the constraint
$(H_c+H_s)|\Psi\>\>=0$ that guarantees that the clock and the system
have (in modulo) equal energy \footnote{Note here that any clock
  Hamiltonian $H_c^\prime$ can be rescaled by a constant $k$, such
  that $(k H_c^\prime+H_s)|\Psi\>\>=0$. In fact, this only results in
  a rescaling of the time parameter (a change of time units) appearing
  in the Schr\"odinger equation. Also, one can add an arbitrary
  additive constant $(H_c^\prime+H_s+k')|\Psi\>\>=0$ without any
  observable consequences.}.

To show this, note first that for time independent $H_s$ the average energy calculated on the system state $\<\psi({t})|H_s|\psi({t})\>$ is independent of $t$ because of energy conservation, and therefore it can also be calculated on $|\Psi\>\>$ giving the same result. In fact we have
\begin{eqnarray}
\<\<\Psi|\openone_c\otimes H_s|\Psi\>\>=\int_{-{T}/2}^{{T}/2}\frac{dt dt'}{T}\<t'|{t}\>\<\psi(t')|H_s|\psi({t})\> \labell{ee} \\\nonumber
=\<\psi(t)|H_s|\psi({t})\>\int_{-{T}/2}^{{T}/2}\frac{dt}{T}=
\<\psi(t)|H_s|\psi({t})\>=\<H_s\>.
\end{eqnarray}
Analogously, we find $(\Delta H_s)_{|\psi({t})\>}=(\Delta H_s)_{|\Psi\>\>}$. We can now use the constraint equation $H_c|\Psi\>\>=-H_s|\Psi\>\>$ to show that the system's average energy is the same as the clock's one
\begin{align}
\<\psi({t})|H_s|\psi({t})\>=\<\<\Psi|H_s|\Psi\>\>=-\<\<\Psi|H_c|\Psi\>\>
\nonumber\\
=-\mbox{Tr}_s[H_c|\Psi\>\>\<\<\Psi|]=-\<H_c\>
\labell{cl112} \;.
\end{align}
This directly implies that $(\Delta H_s)_{|\Psi\>\>}=(\Delta H_c)_{|\Psi\>\>}$.

If $[\Pi,H_s]=0$, we can condition the energy of the system on the event, and define the conditioned energy observable as $\Pi H_s\Pi$. Its expectation value is
\begin{eqnarray}
&&  \<E_{ev}\>=\int dE\: E\: p(E|\Pi)=\int
  dE\:E\:\frac{p(E,\Pi)}{p(\Pi)}\labell{me}\\&&\nonumber
=\<\psi({t})|\Pi H_s\Pi|\psi({t})\>/p(\Pi)=\<\psi({t})|\Pi H_s\Pi|\psi({t})\>\alpha T \;,
\end{eqnarray}
where $p(E|\Pi)$ and $p(E,\Pi)$ are the conditional and the joint probability distributions for the energy $E$ and the event $\Pi$ happening. Analogously, one can calculate the variance
\begin{align}
\Delta E_{ev}^2=\alpha {T}\<\Pi H_s^2\Pi\>-(\alpha {T})^2\<\Pi H_s\Pi\>^2
\labell{me0}\;.
\end{align}
As shown before, both the expectation value $\<E_{ev}\>$ and the RMSE $\Delta E_{ev}$ can be calculated on the state $|\Psi\>\>$ obtaining the same result
\begin{eqnarray}
&&\<\<\Psi|E_{ev}|\Psi\>\>=
\int\frac{dtdt'}{{T}}\<t'|{t}\>\<\psi(t')|\Pi
  H_s\Pi|\psi({t})\>/p(\Pi)
\nonumber\\&&=
\<\psi({t})|\Pi H_s\Pi|\psi({t})\>/p(\Pi)\int_{-{T}/2}^{{T}/2}\dfrac{dt}{T} 
\labell{me2}\;,
\end{eqnarray}
where the last integral is equal to one. A similar relation holds for computing $(\Delta E_{ev})_{|\Psi\>\>}$. Again, thanks to the constraint $H_s|\Psi\>\>=-H_c|\Psi\>\>$, the above relations written in terms of the system's energy $H_s$ can be also written equivalently in terms of the clock energy $H_c$ using
\begin{align}
\<\<\Psi|\openone_c\otimes\Pi H_s\Pi|\Psi\>\>=-\<\<\Psi|H_c\otimes\Pi|\Psi\>\>
\labell{cl}\;,
\end{align}
where $[H_s,\Pi]=0$ was used.

If, instead, $[H_s,\Pi]\neq 0$, we can use Eq.~\eqref{cl112} to argue that a conditioned energy can be obtained by looking at the clock energy, conditioned on the event having happened on the system. We obtain $E_{ev}=-\<\<\Psi|H_c\otimes\Pi|\Psi\>\>\alpha T$ and $\Delta E^2_{ev}=\alpha {T}\<(H_c\otimes\Pi)^2\>-(\alpha{T}\<H_c\otimes\Pi\>)^2$, where the $\alpha T$ terms come from the Bayes rule, as in Eqs.~\eqref{me}, \eqref{me0}.

In both cases just considered, defining $H_\pi\equiv H_c\otimes\Pi$ we have
\begin{align}
\Delta E^2_{ev}=
\alpha {T}\<\<\Psi|H_\pi|\Psi\>\>-(\alpha {T})^2
\<\<\Psi|H_\pi|\Psi\>\>^2
\labell{dele}\;.
\end{align}
As for the conditioned time, also the conditioned energy satisfies $\Delta E_{ev}^2=\alpha {T}\Delta H_\pi^2$ because a shift $E_0$ in the average value of $E_{ev}$ corresponds to a shift $\alpha {T} E_0$ of the average value of $H_\pi$.

\textit{Uncertainty relation.--} 
We can now derive Eq.~\eqref{teunc} as
\begin{align}
  \Delta {t}^2_{ev}\Delta E^2_{ev} &= (\alpha {T})^2\Delta {T}^2_\pi
\Delta H^2_\pi \geqslant (\alpha {T})^2 |\<[{T}_\pi,H_\pi]\>|^2/4\nonumber\\
&=(\alpha{T})^2\hbar^2\<\Pi\>^2/4=\hbar^2/4 
\labell{teuncproof}\;,
\end{align}
where the inequality follows from the Robertson uncertainty relation for ${T}_\pi$ and $H_\pi$ calculated on $|\Psi\>\>$, and the second row is obtained from $[{T}_\pi,H_\pi]=[{T}_c,H_c]\otimes\Pi^2=i\hbar\openone_c\otimes\Pi$.

In a similar way, we can also find an uncertainty relation for the unconditioned system energy:
\begin{align}
\Delta {t}_{ev}^2\Delta H_s^2 &=  \alpha {T}\Delta {T}_\pi^2\Delta H_s^2 \notag\\ 
&\geqslant\dfrac{\alpha{T}}{4}\left|\<\<\Psi|[{T}_\pi,H_s]|\Psi\>\>\right|^2
\labell{teuncdim}\;
\end{align}
where the inequality sign comes again from the Robertson uncertainty relation between ${T}_\pi$ and $H_s$, with both variances calculated on $|\Psi\>\>$.
Using the constraint equation and the condition $[{T}_c,H_c]=i\hbar$, we find
\begin{align}
\<[{T}_\pi,H_s]\> &= \<-({T}_cH_c\otimes\Pi)+(H_c{T}_c\otimes\Pi)\> \\
&= -\<[{T}_c,H_c]\otimes\Pi\>
=-{i\hbar}/{\alpha {T}} \;,
\labell{teuncdim2}\;
\end{align}
which, joined together with \eqref{teuncdim}, gives
\begin{align}
\Delta {t}_{ev}\Delta H_s \geqslant \dfrac{\hbar}{2} \sqrt{P_{ev}}
\labell{teunc2}\;,
\end{align}
where $P_{ev}=p(\Pi)$ is the overall probability that the event happens. Unfortunately, it appears that this last relation is always trivial: if the event does not happen an infinite number of times, then in the limit ${T}\to\infty$ we have $P_{ev}\to 0$, if it does happen an infinite number of times, then clearly $\Delta{t}_{ev}\to\infty$. In both cases, the inequality \eqref{teunc2} is satisfied trivially.

\textit{Examples.--} 
As a first example of the above inequalities, consider a photon with spectral amplitude $\varphi(\omega)$, namely the state $|\psi\>=\int \tfrac{d\omega}{2\pi}\varphi(\omega)|1\>_\omega$, where $|1\>_\omega={a_{\vec k}}^\dag|0\>$ is a single photon at frequency $\omega=|\vec k|c$ (assuming a mode with fixed spatial direction $\vec k/|\vec k|$). The free evolution of $|\psi\>$ governed by the electromagnetic field Hamiltonian $H_s=\int d\omega\hbar\omega a^\dag_\omega a_\omega$ induces the phase shift $\varphi(\omega)\to\varphi(\omega)\:e^{-i\omega {t}}$, so that the timeless state reads
\begin{align}
|\Psi\>\>=\int \frac{dt}{\sqrt{T}}|{t}\>\int
\frac{d\omega}{\sqrt{2\pi}}\:\varphi(\omega)e^{-i\omega {t}}|1\>_\omega,
\labell{st11}\;
\end{align}
The phase factor appearing in Eq.~\eqref{st11} is equivalent to a translation $z$ along the propagation direction $\vec k/|\vec k|$: the photon wavepacket propagates at a speed $c$ determined by the Klein-Gordon equation $\square(a_{\vec k}e^{-i(\omega {t}-\vec k\cdot\vec x)}+h.c.)=0$ for all $\vec k$.  Consider now a screen placed perpendicularly to the propagation at the position $z_0$. The projector associated to the detection of a photon by the screen is
\begin{align}
\Pi_{{z}_0}=|1\>_{{t}_0}\<1|\equiv\int d\omega d\omega'e^{it_0(\omega'-\omega)}(|1\>_\omega)(_{\omega'}\<1|)
\labell{proj}\;,
\end{align}
with ${t}_0=z_0/c$. Then, as expected, the probability amplitude that the photon arrives at time ${t}$ is the Fourier transform $\tilde\varphi({t})\propto\int d\omega\:e^{-i\omega{t}}\varphi(\omega)$. In fact we have
\begin{align}
p({t}|\Pi)=\<\<\Psi|(|{t}\>\<{t}|\otimes\Pi_{{z}_0}|\Psi\>\>/p(\Pi_{{z}_0})=
|\tilde\varphi({t}-{t}_0)|^2
\labell{pp}\;,
\end{align}
assuming $_{\omega'}\<1|1\>_\omega=\delta(\omega-\omega')$ (which follows from the Dirac-delta commutators of the $a_{\vec k}$, restricting to positive energies) and with the normalization $\int d\omega|\varphi(\omega)|^2/2\pi=\int dt|\tilde\varphi({t})|^2=1$. Note that the regularization factor $T$ does not appear in Eq.~\eqref{pp} thanks to $p(\Pi)=1/T$.  Then, $\Delta {t}_{ev}$ is clearly the width $\Delta t$ of the probability distribution $|\tilde\varphi({t})|^2$. The energy of the photon conditioned on having arrived at position $z_0$ can be calculated as
\begin{eqnarray}
&&\<E_{ev}\>=-\<\<\Psi|H_c\otimes\Pi_{z_0}|\Psi\>\>/p(\Pi_{z_0})=\labell{enpho}\;\\
&&\!\!-\!\int\! \dfrac{dt\:dt'}{2\pi}\<t'|H_c|{t}\>\tilde\varphi^*(t'-{t}_0)\tilde\varphi(t-{t}_0) =\!\int\! \dfrac{dp}{2\pi}\:\hbar p|\varphi(p)|^2 .\nonumber
\end{eqnarray}
Similarly, $\Delta E_{ev}/\hbar$ can be shown to coincide with the width $\Delta\omega$ of the spectrum $|\varphi(\omega)|^2$. The Parseval inequalities, expressing a lower bound to the time--bandwidth product, imply that the widths $\Delta t$ and $\Delta\omega$ of the probability distributions $|\tilde\varphi({t})|^2$ and $|\varphi({\omega})|^2/2\pi$ satisfy $\Delta {t}\Delta\omega\geqslant 1/2$. This then directly implies the validity of inequality \eqref{teunc} for this example.

Another example is the case in which the event consists in observing a
photon of frequency $\omega_0$, namely the projector
$\Pi_{\omega_0}=|1\>_{\omega_0}\<1|$. In this case, we find
$p(\Pi)=|\varphi(\omega_0)|^2$. Therefore, if $\omega_0$ is in the
support of $\varphi$, $p({t}|\Pi_{\omega_0})=1/T$ is a constant (where
the regularization $T$ cannot be eliminated). This implies that
$\Delta {t}_{ev}=\infty$, while $\Delta E_{ev}$ is finite (the width
of $\varphi$), so inequality \eqref{teunc} is trivially
satisfied.

\textit{Conclusion.--} 
Time-energy uncertainty relations commonly found in the literature relate the energy spread of a system to the length of a time interval, and not to the uncertainty in the measurement of time due to quantum noise. In this sense, they are better understood as quantum speed limits in the time evolution of a state. In contrast, we presented here uncertainty relations that relate the uncertainty in the quantum measurement of the time at which a quantum event happens on a system to its energy uncertainty. 

Our results are of foundational interest, as they clarify the connection between time intervals and time measurements: if the system takes an interval at least $\Delta t$ to evolve to an orthogonal state, then an event cannot even be defined on a shorter time scale (in accordance to quantum speed limits). This means that the measurement outcome of when the event happens will have at least that quantum uncertainty associated to it.

In addition, our results might also be of practical relevance, and find verification in state-of-the-art experiments \cite{marco1,marco2}.

\end{document}